# Practical Post-Quantum Cryptography for Bandwidth Constrained or Non-Terrestrial Networks, and Power Constrained Devices

Elliot Eichen, Sylvia Llosa, Yueqi Chen, and Sangtae Ha
Computer Science, University of Colorado Boulder
Boulder, CO, USA
{elliot.eichen, sylvia.llosa, yueqi.chen, sangtae.ha}@colorado.edu

***Abstract*—Post-quantum (PQ) cryptographic algorithms, particularly authentication algorithms, are more complex than classical cryptographic algorithms and require significantly larger certificates, signatures, and keys. Establishing a PQ-secure network connection can thus incur increased bandwidth consumption, memory usage, computation time, and energy per transmitted bit. These costs are especially severe in Non-Terrestrial Networks (NTNs), where long propagation delays, intermittent connectivity, constrained link budgets, satellite handovers, limited terminal resources, and bandwidth-constrained satellite-to-ground links amplify the overhead associated with certificate-based PQ authentication. As a result, critical applications using NTNs, such as key rotation and key management, may be difficult to implement with a reasonable probability of a successful cryptographic handshake, or with NIST PQ Security level greater than Level 1. Similar constraints also affect low-power IoT devices and bandwidth, or energy-limited terrestrial networks. PQ authentication can restrict low-power IoT applications to Level 1 security, while ambient-powered endpoints may be unable to support even Level 1 operation under realistic network conditions. This paper investigates an alternative cryptographic framework that replaces PQ digital certificates with shared secret keys (SSKs). The proposed framework leverages shared secret-key ecosystems that do not depend on asymmetric algorithms for key distribution, such as 5G/6G, and combines a Key Distribution Center (KDC) (such as Kerberos, or an equivalent key management service) with a pre-shared-key (PSK) PQ handshake, such as DTLS 1.3-PSK with an ephemeral PQ key establishment primitive (e.g., ML-KEM). Compared with conventional certificate-based PQ authentication using ML-DSA or Falcon in conjunction with ML-KEM, the proposed approach substantially reduces handshake bandwidth, endpoint RAM requirements, computation time, and energy consumption while preserving PQ-secure Authenticated Encryption with Associated Data (AEAD), including forward secrecy and replay resistance. Representative applications include NTN-based key rotation and management, Uncrewed Aerial Vehicles (UAVs) command-and-control systems, embedded medical sensors, and supply-chain monitoring/asset-tracking platforms.***

***Keywords*—post-quantum cryptography, post-quantum authentication, key rotation, shared secret keys, IoT, ISAC, non-terrestrial-networks***

## I. Introduction

Fault-tolerant quantum computers pose a significant threat to conventional asymmetric cryptographic systems, including RSA and elliptic-curve cryptography (ECC) [1], [2]. Recent studies project that such quantum computers may emerge within the next decade [3], [4] enabling practical attacks against widely used asymmetric public-key and private-key cryptographic systems. The threat is also immediate because encrypted data can be collected today and decrypted retrospectively once large-scale quantum computers become available; this vulnerability is commonly referred to as "harvest now, decrypt later."

The cryptographic community's primary response has been the development of post-quantum (PQ) asymmetric algorithms to replace existing asymmetric algorithms for authentication and key establishment [5]. However, certificate-based PQ authentication introduces larger keys, signatures, and certificates than conventional algorithms [6]. This results in PQ cryptographic handshakes with significant bandwidth, memory, computational, and energy burdens on constrained systems and low-power devices [7]. In practical systems, PQ certificate-based authentication, rather than PQ key establishment, is the dominant source of increased packet overhead, handshake latency, memory utilization, and energy consumption [8].

TABLE 1: NETWORK AND DEVICE METRICS FOR NON-PQ ELLIPTIC CURVE AND PQ ML DSA/KEM SECURITY. (SEE APPX. I FOR DETAILS)

| Authentication Algorithm | P-256 ECDSA (not PQ safe) | P-521 ECDSA (not PQ safe) | ML-DSA-44 ML-KEM-512 | ML-DSA-87 ML-KEM-1024 |
|---|---|---|---|---|
| PQ Level | 1 (equiv) | 5 (equiv) | 1 | 5 |
| Crypto RAM [kbytes] | 12-24 | 16-32 | 32-80 | 64-150 |
| Crypto Payload [bytes] | 120-150 | 123 | 5300 | 10500 |
| # Packets (NB-IoT) | 1 | 2 | 46 | 88 |
| Real World # Packets | 8 | 15 | 334 | 638 |
| Tx+Rx Energy [mJoules] | 256 | 480 | 10688 | 20416 |
| % Battery/Handshake | 0.013% | 0.020% | 0.54% | 1.03% |
| Ambient Powering ? | Possible | Possible | Difficult | Not Feasible |
| SoC Compute Time [ms] | 30 | 40 | 55 | 175 |

These limitations are especially problematic for bandwidth-constrained networks and applications, including Non-Terrestrial Networks (NTNs) that support key rotation and key management systems, as well as for power-constrained networks and devices, such as UAVs using low-power wide-area networks (LPWANs). These applications, and many others that rely on NTNs, LPWANs, or low-power devices, often have difficulty in achieving NIST PQ Security Level 1 (equivalent to AES-128 for conventional non-PQ security) [9]. Many ambient-powered device applications cannot support even PQ Level 1 security [9]. Although approaches such as pre-installed certificates or reduced handshake frequency can mitigate some PQ-authentication costs, many operational environments, including zero-trust architectures, roaming devices, intermittently connected systems, and hostile or contested environments, cannot rely on long-lived sessions or infrequent authentication exchanges. Table 1 compares network and device metrics for non-PQ and PQ asymmetric algorithms transmitted over a constrained NB-IoT network using a typical low-power System-on-Chip (SoC).

To address the constraints of certificate-based PQ authentication, this paper proposes using existing symmetric-key distribution ecosystems to enable scalable PQ authentication. The main idea is to replace asymmetric endpoint authentication with shared-secret-key authentication while preserving post-quantum ephemeral session-key establishment through ML-KEM. The proposed architecture combines a Kerberos KDC (Key Distribution Center), or equivalent KDC, with shared secret keys (SSKs) and a pre-shared-key (PSK) cryptographic handshake that uses an ephemeral PQ key-establishment algorithm to support forward secrecy. This approach can reduce server-side authentication bottlenecks [10], enable PQ Security Levels of 3 or greater on moderate-resource SoCs with sufficient memory, and support PQ Level 1 operation on some ambient-powered devices.

Symmetric cryptographic primitives are generally believed to remain resistant to known quantum attacks when appropriately sized SSKs. SSK-based authentication is used today by enterprise-scale systems such as Microsoft Exchange, AWS Single Sign-On, and Oracle databases. The central challenge is not the cryptographic primitives, but scalable key distribution and lifecycle management. Historically, that challenge has led many deployments to rely on asymmetric public-key infrastructure (PKI) for secure key distribution. This work addresses that obstacle by leveraging existing SSK ecosystems that do not use PKI to derive and distribute new SSKs, and Trusted Platform Modules (TPMs) for key storage. Examples include 3GPP-compliant mobile and wireline networks, where long-term SSKs are securely stored by the network database and within an endpoint's SIM or iSIM [11], and digital identity infrastructures, where long-term SSKs are stored in digital ID cards [12].

The proposed architecture combines three components: (1) an existing non-PKI SSK ecosystem for secure key derivation and distribution, (2) a Key Distribution Center for scalable authentication and session-key management, and (3) a pre-shared-key cryptographic handshake, such as DTLS-PSK [11], CoAP [12], or TLS-PSK [13], that employs an ephemeral PQ key-establishment algorithm such as ML-KEM [14]. In one practical implementation, a lightweight private 5G core and a 5G endpoint separately derive new SSKs and securely transfer them to a Kerberos KDC [15]. Kerberos then enables endpoints and application servers to mutually authenticate and derive session-specific shared keys without requiring a full mesh of pre-shared keys between devices and servers. These derived keys subsequently serve as the PSKs for the cryptographic handshake.

A private 5G/6G infrastructure is particularly attractive because it offers a mature, widely deployed, and extensively analyzed security infrastructure, along with existing support for hierarchical key derivation and secure credential provisioning. The proposed approach uses only a small subset of 5G core functionality (specifically, authentication and key management) and remains independent of the underlying transport network. Consequently, the methodology can support many constrained transport protocols, including NB-IoT [16], LoRaWAN [17], Bluetooth Low Energy (BLE) [18], and 6LoWPAN [19]. In addition, 3GPP Release 15 introduces network key derivation mechanisms [20] that accelerate the practical deployment of this architecture.

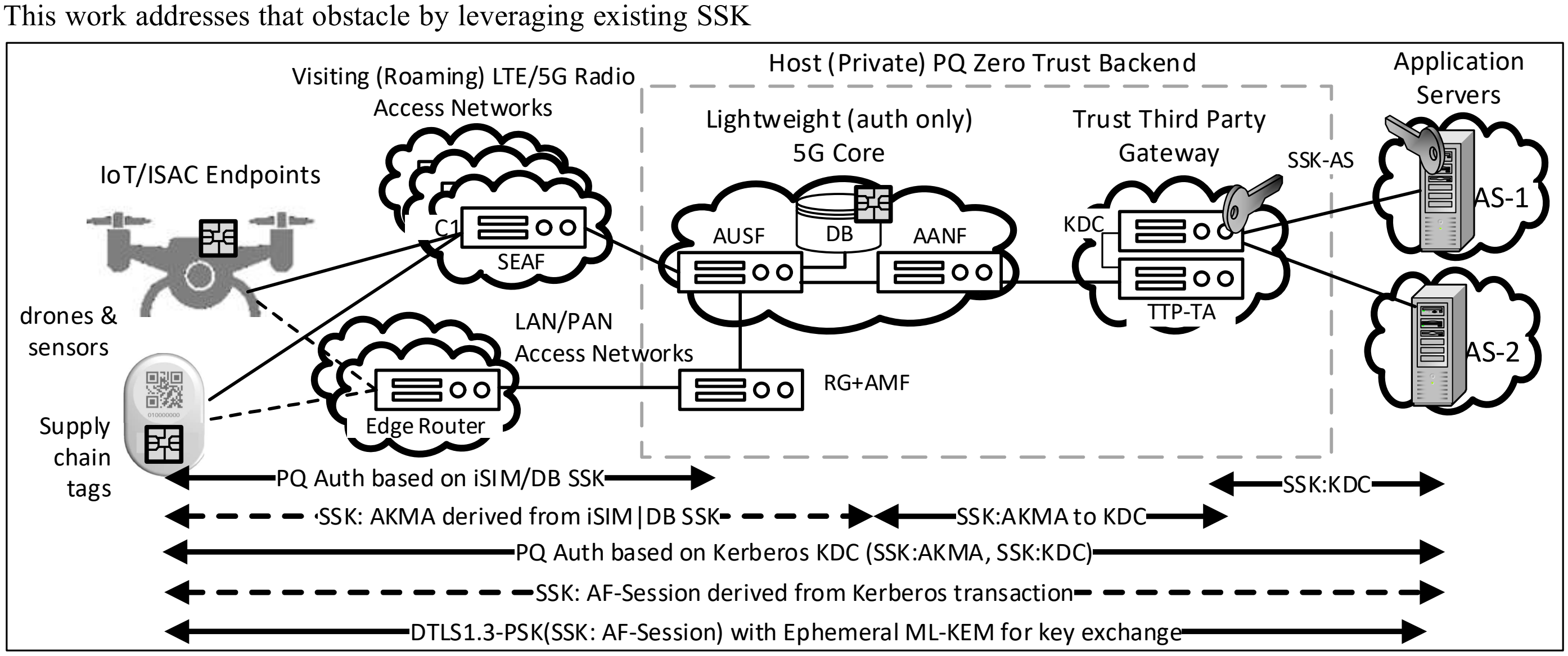

Fig. 1: PQ end-to-end secure network with forward secrecy based on PSK authentication with Kerberos KDC and ephemeral PQ key establishment (ML-KEM).

### PROPOSED ARCHITECTURE: END-TO-END PQ SECURITY WITH 5G AND KERBEROS

Fig. 1 illustrates one implementation of the proposed architecture. An IoT endpoint first mutually authenticates with a host private 5G network using the long-term SSK shared between the endpoint's integrated SIM (iSIM) and the network database. When the endpoint requests registration with a 3GPP network, the host network database generates an authentication vector based on the long-term SSK. The endpoint decrypts components of the authentication vector and returns the appropriate response, enabling it to register via the host network. After this exchange, the endpoint and network independently derive a new SSK based on data from the registration using a one-way key derivation function [21]. The new SSK is then transferred to the KDC through a PQ-secure transfer mechanism. Kerberos then enables the endpoint and application server to mutually authenticate and derive a session-specific key that serves as the PSK for a DTLS 1.3-PSK session with ephemeral ML-KEM key establishment.

TABLE 2: COMPARISON OF CONVENTIONAL ML-DSA WITH ML-KEM TO DTLS-PSK WITH EPHEMERAL ML-KEM OVER NB-IOT (APPENDIX 1)

| Authentication Algorithm | ML-DSA-44 ML-KEM-512 | ML-DSA-87 ML-KEM-1024 | DTLS-PSK (128) ML-KEM-512 | DTLS-PSK (256) ML-KEM-1024 |
|---|---|---|---|---|
| PQ Level | 1 | 5 | 1 | 5 |
| Crypto RAM [kbytes] | 32-80 | 64-150 | 16-40 | 24-64 |
| Crypto Payload [bytes] | 5300 | 10500 | 1568 | 3136 |
| # Packets (NB-IoT) | 46 | 88 | 14 | 27 |
| Real World # Packets | 334 | 638 | 112 | 216 |
| Tx+Rx Energy [mJoules] | 10688 | 20416 | 3584 | 6912 |
| % Battery/Handshake | 0.54% | 1.03% | 0.16% | 0.32% |
| Ambient Powering ? | Difficult | Not Feasible | Feasible | Difficult |
| SoC Compute Time [ms] | 55 | 175 | 8 | 15 |

Table II compares conventional ML-DSA/ML-KEM with DTLS transport against the proposed Kerberos + DTLS-PSK + ephemeral ML-KEM system. Under the same assumptions (constrained NB-IoT network) used for Table I and described in Appendix I, the proposed architecture requires approximately 70% less bandwidth, one-half the RAM, one-third the handshake energy, and one-tenth the compute time.

Fig. 2 provides a simplified sequence diagram and key hierarchy for establishing an end-to-end PQ-secure communication link between an endpoint and an application server (AS). The architecture is designed to provide authenticated encryption, forward secrecy, integrity protection, and replay resistance. The system architecture has four logical phases: (1) initial trust relationships, (2) 5G endpoint authentication and Authentication and Key Management for Applications (AKMA) key derivation, (3) Kerberos-based endpoint-to-application-server mutual authentication, and (4) a PQ-secure DTLS-PSK handshake with ephemeral ML-KEM key establishment for encryption.

**Step 0: Initial Conditions (Trust Relationships).** The initial system state assumes the existence of several long-term trust relationships. First, the endpoint's integrated SIM (or iSIM [22]) and the host 5G network share a long-term symmetric secret key, $K_{5G}$, consistent with conventional 3GPP authentication procedures. Second, each application server is assumed to be a registered Kerberos client that possesses a long-term shared secret, $K_{AS}$, with the KDC. In addition, all links within the Zero Trust DMZ are assumed to be PQ-secure at the appropriate security level, using PQ asymmetric algorithms, quantum links, SSKs, or a combination of these mechanisms.

**Step 1: 5G Endpoint Authentication and AKMA Key Derivation.** The first operational phase consists of 5G endpoint registration and authentication using the conventional 3GPP authentication framework [12]. Note that 5G registration supports roaming into a foreign carrier's Radio Access Network (RAN), tunneling through a broadband network rather than a RAN, or transport through an IP based (IPX [23]) interconnection network. None of these access mechanisms compromises the long-term secret key between the endpoint and the host 5G network database or the session key for device registration. The proposed architecture also leverages 3GPP Authentication and Key Management for Applications (AKMA) [24] to derive an application-specific shared secret key, $K_{AKMA}$.

**Step 2: Kerberos-Based Endpoint-to-Application-Server Mutual Authentication.** To authenticate with an application server, the endpoint initiates a request through the trusted third-party transfer agent (TTP-TA) that triggers the retrieval of $K_{AKMA}$ from the AKMA Anchor Function (AANF). The transferred $K_{AKMA}$ may be used directly as the Kerberos shared secret for the 5G endpoint, or as input to a subsequent key-derivation process that generates a dedicated Kerberos authentication key, $K_{KDC}$. Once the Kerberos shared secret key for the endpoint has been established, conventional Kerberos mutual authentication between the endpoint and the AS proceeds normally [15]: the endpoint obtains a ticket-granting ticket, uses it to obtain an authentication ticket, and presents that ticket to the AS.

**Step 3: PQ-Secure DTLS-PSK Authenticated Encryption.** After completing Kerberos authentication, the endpoint and AS each derive the shared secret key $K_{AS\text{-}SESSION}$ from material in the Kerberos ticket in Step 2. A conventional DTLS 1.3-PSK handshake can then proceed using $K_{AS\text{-}SESSION}$ as the PSK and an ephemeral PQ key-establishment algorithm such as ML-KEM.

The result of Steps 0–3 is an end-to-end PQ-secure Authenticated Encryption with Associated Data (AEAD) communications channel between the endpoint and the application server. The proposed architecture provides mutual authentication, authenticated encryption, message integrity, replay protection, scalable key management, and forward secrecy.

## II. EXAMPLE USE CASES

### A. Supply Chain Monitoring and Asset Tracking

Supply chain monitoring and asset tracking are representative IoT applications that require PQ security, low device cost, long operational lifetimes, and low-power wireless connectivity.

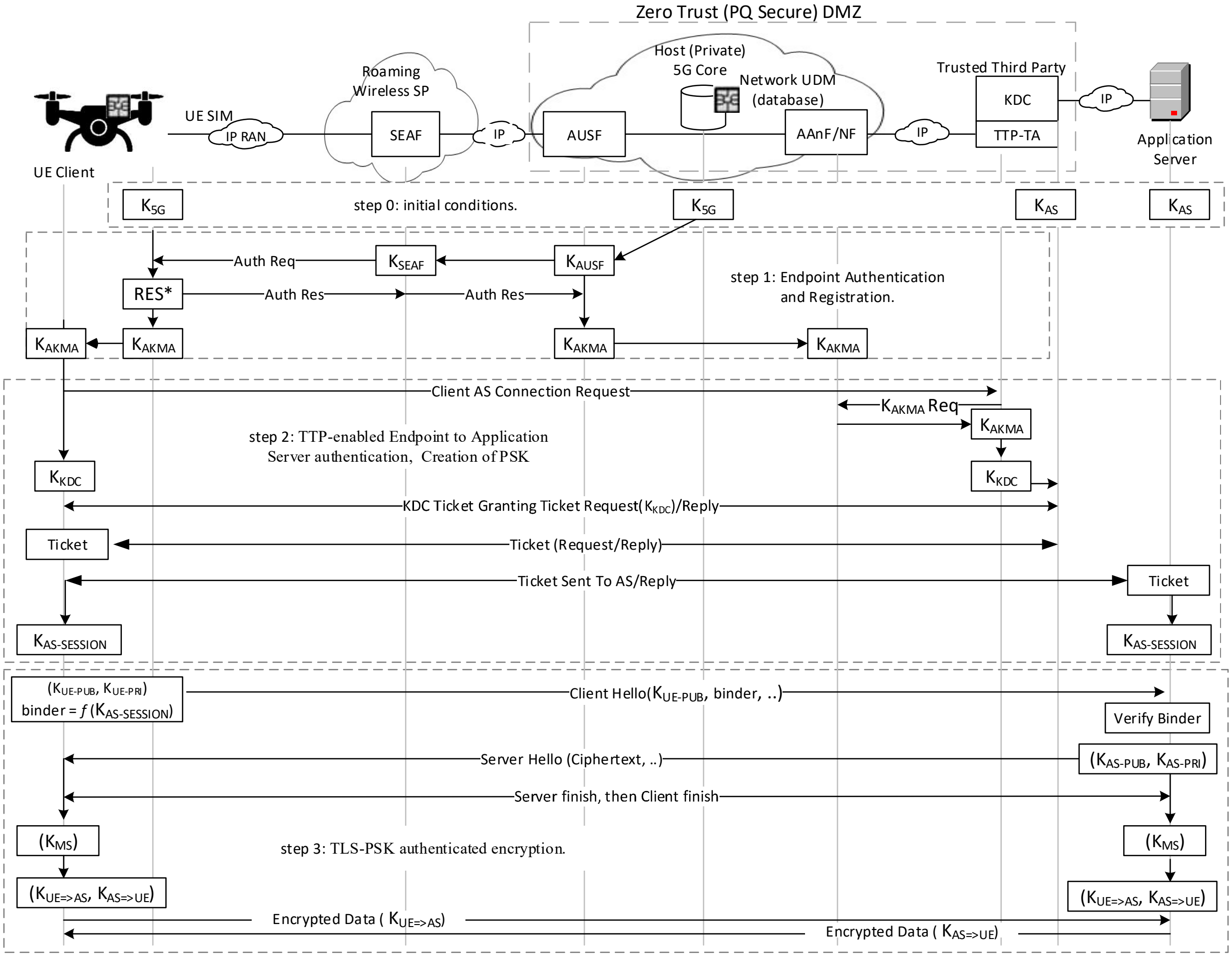

Fig. 2: Simplified ladder diagram and key hierarchy for enabling a PQ-secure authenticated-encryption link between a 5G endpoint and an Application Server using shared secret keys for authentication (and thus without requiring digital signatures). Note that there are many other 5G architectures that can use broadband connectivity and/or IP transport exchanges, or other registration architectures. Note that there are also other Shared Secret Key ecosystems that could be used instead of 3GPP 5G.

Typical deployment environments include LPWANs and LPPANs (Low Power Personal Area Networks, such as Bluetooth Low Energy or 6LoWPAN). Fig. 3 illustrates two representative commercial tracking devices for monitoring package location, temperature, humidity, shock, and light exposure. The smaller device, or tag, uses BLE connectivity and operates opportunistically by transmitting sensor data when a trusted BLE gateway is detected. This device uses an ARM Cortex-M33 SoC powered by a non-rechargeable CR2032 lithium coin-cell battery. The larger device uses LTE Cat-M1 connectivity based on an ARM Cortex-A7 platform, requires frequent recharging and provides continuous wide-area telemetry, including GPS location.

Conventional certificate-based PQ authentication would impose substantial bandwidth, memory, and energy overhead on both device classes. The methodology proposed in this paper would enable both platforms to support PQ-secure authenticated encryption while remaining within practical power and hardware constraints.

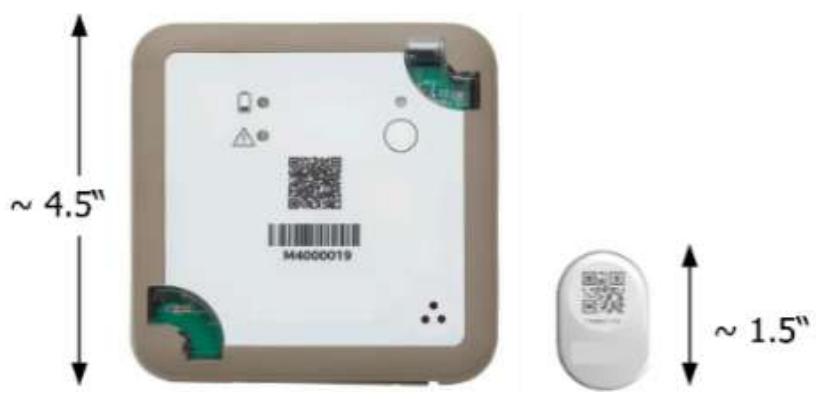

Fig. 3: Example Supply Chain Asset Tracking Devices

## B. *UAV Detection and Control*

PQ security for detecting and controlling UAVs, especially around critical infrastructure such as airports and government facilities, is required to prevent data and application compromise and potential device hijacking. While first-generation Integrated Sensing and Communications (ISAC) UAV tracking may not require LPWANs, communication with and potential control of UAVs do. Deep ISAC integration of UAV tracking with 5G/6G will also benefit from UAVs being simultaneously connected to the same base station [25].

### C. Embedded Medical Sensors

Medical devices embedded under the skin, also called subdermal implants, can be used to monitor vital signs (such as atrial fibrillation or hemodynamic status), track glucose levels, or control medication delivery. Implanted devices have very restrictive battery power requirements because inductive charging is not feasible due to skin-effect heating. Devices designed for home monitoring often use Bluetooth Low Energy Personal Area Networks, while institutional settings, such as hospital intensive care units and nursing homes, may require more sophisticated Personal Area Networks such as 6LoWPAN. In both cases, PQ-secure networks and devices may be required to support future health-sector privacy and cybersecurity requirements. [26].

### D. Non-Terrestrial Networks and Key Rotation Systems

The proposed SSK-based authentication methodology is also well-suited for non-terrestrial networks (NTNs) [27] [28] including low Earth orbit (LEO) satellite systems that host key-rotation and key-management systems. Communications over these links often experience high packet loss due to atmospheric conditions, link intermittency, handoffs, and congestion; these issues are exacerbated by the long round-trip times (RTTs) for LEO satellites. The application of symmetric PSK algorithms for PQ key-rotation systems is therefore important for three reasons:

1. **Reliability over Lossy Satellite Links**: The probability $P$ of a successful PQ handshake is given by $P \sim (1-p)^N$ where p is the packet loss probability, and N is the number of packets in the handshake protocol. Because the probability of success is exponentially dependent on the number of packets, even small reductions in packet counts can dramatically change the probability of success. For example, with $p = 10\%$ (which is reasonable for an average NTN NB-IoT link), reducing N from 46 packets (ML-DSA) to 14 packets (PSK based algorithms) for PQ Level 1 (Table 2) increases the handshake success rate to 23% for the PSK algorithm from 0.8% for the asymmetric algorithm. (Note that this model is first order in that it assumes that the probability of losing any packet is independent of losing any other packet; it may therefore overestimate handshake success probability in wireless environments that experience burst losses or correlated fading. However, it may also underestimate the overall probability because it assumes there is no recovery from lost packets.)
2. **Long Handshake Times**: Long PQ handshakes significantly increase exposure to beam handoffs, intermittent visibility, and satellite mobility events, potentially spanning multiple beam or satellite transitions [30]. For $p = 10\%$ (as above), the expected number of handshake attempts is ~5 for the PSK based handshake, and ~125 for the classical PQ algorithm-based handshake (assuming a complete handshake restart if a packet is lost). This leads to handshake completion times of ~7 sec and ~6 min for the PSK and classical PQ handshakes, respectively. With a 10–15 second window to complete a handshake to a single satellite (depending on antenna and satellite characteristics), it is clear that the PSK-based handshakes are feasible depending on antenna and satellite characteristics, but that conventional certificate-based PQ handshakes significantly increase exposure to beam handoffs, intermittent visibility, and satellite mobility events, potentially spanning multiple beam or satellite transitions. Reducing total handshake latency and retransmission exposure can significantly simplify the key-management process.
3. **Lifecycle Key Management**: A central challenge in scalable symmetric-key authentication is secure key distribution and lifecycle management [29]. This work addresses that challenge by potentially reusing the same mature commercial off-the-shelf 5G/6G technology used for PQ authentication to provision, store, rotate, and manage keys.

For these reasons, the authors believe that PQ security using symmetric PSK algorithms and 5G/6G cores is an important use case worth exploring.

ACKNOWLEDGMENT

The author gratefully acknowledges comments and suggestions from Professors Kevin Gifford and Dirk Grunwald, University of Colorado Boulder.

## APPENDIX 1

The comparisons in Tables 1 and 2 use a common constrained-link model. Packetization assumes IPv6 and UDP/DTLS over 200-byte NB-IoT packets, with 80 bytes of overhead per packet. ECDSA calculations assume compressed public keys and Distinguished Encoding Rules (DER) for signatures, while ML-DSA signature sizes are treated as fixed. The PQ payload model pairs ML-KEM-512 with ML-DSA-44 and ML-KEM-1024 with ML-DSA-87, and cryptographic payloads calculated from the corresponding ML-KEM and ML-DSA parameter sets.

Device-side metrics assume an ARM Cortex-M33 clocked at 133 MHz. NB-IoT throughput is modeled at 20 kb/s, although practical values can range from roughly 10–60 kb/s. At 20 kb/s, a 200-byte NB-IoT packet corresponds to approximately 80 ms of transmission time.

The packet and energy handshake estimates also include real-world transport and radio effects. Moving from NB-IoT payload packets to real-world packets incorporates protocol-level inefficiencies, including partially filled DTLS fragments, and radio-layer reliability mechanisms such as repeated transmissions and lower coding rates. The model uses a factor of 1.25 for protocol-layer inefficiency and a (typical) factor of 6 for weak-signal radio reliability. Energy calculations assume a CR2032 battery providing 220 mAh at 3.3 V, or approximately 2375 J, with a transmit current of 180 mA and a receive current of 60 mA. Under these assumptions, per-packet energy is approximately 48 mJ for transmission and 16 mJ for reception. NB-IoT modems may "silently" repeat packets, so the physical/radio-layer retransmission factor can increase from 2 to 128 depending on link quality and retransmission rates